\documentclass[12pt]{amsart}
\usepackage{amsmath,amscd,amsthm}

\theoremstyle{plain}

\newtheorem*{thm*}{Theorem}

\newtheorem*{conj*}{Conjecture}

\theoremstyle{definition}
\newtheorem*{defn*}{Definition}

\newtheorem*{rems*}{Remarks}
\newtheorem*{proof*}{Proof}

\numberwithin{equation}{section}
\numberwithin{thm}{section}

\newcommand{\cK}{{\mathcal K}}
\newcommand{\cU}{{\mathcal U}}

\newcommand{\cH}{{\mathcal H}}

\newcommand{\cJ}{{\mathcal J}}
\newcommand{\cE}{{\mathcal E}}
\newcommand{\cD}{{\mathcal D}}
\newcommand{\cQ}{{\mathcal Q}}
\newcommand{\CA}{{\mathcal A}}
\newcommand{\cB}{{\mathcal B}}
\newcommand{\CK}{{\mathcal K}}
\newcommand{\CH}{{\mathcal H}}

\newcommand{\ZZ}{\mathbb{Z}}

\def\cA{{\mathcal A}}

\newcommand{\nc}{\newcommand}
\nc{\nt}{\newtheorem}
\nc{\gf}[2]{\genfrac{}{}{0pt}{}{#1}{#2}}
\nc{\mb}[1]{{\mbox{$ #1 $}}}
\nc{\real}{{\mathbb R}}
\nc{\comp}{{\mathbb C}}
\nc{\ints}{{\mathbb Z}}
\nc{\Ltoo}{\mb{L^2({\mathbf H})}}
\nc{\rtoo}{\mb{{\mathbf R}^2}}
\nc{\slr}{{\mathbf {SL}}(2,\real)}
\nc{\slz}{{\mathbf {SL}}(2,\ints)}
\nc{\su}{{\mathbf {SU}}(1,1)}
\nc{\so}{{\mathbf {SO}}}
\nc{\hyp}{{\mathbb H}}
\nc{\disc}{{\mathbf D}}
\nc{\torus}{{\mathbb T}}

\nc{\ca}{{\mathcal A}}
\nc{\cag}{{{\mathcal A}^\Gamma}}
\nc{\cg}{{\mathcal G}}
\nc{\chh}{{\mathcal H}}
\nc{\ck}{{\mathcal B}}
\nc{\cl}{{\mathcal L}}
\nc{\cm}{{\mathcal M}}
\nc{\cs}{{\mathcal S}}
\nc{\cz}{{\mathcal Z}}
\nc{\G}{{\mathcal G}}
\nc{\cF}{{\mathcal F}}
\nc{\Sc}{{\mathcal S}}
\nc{\E}{{\mathcal E}}

\def\cH{{\mathcal H}}
\def\cK{{\mathcal K}}
\def\cA{{\mathcal A}}
\def\cU{{\mathcal U}}
\def\cE{{\mathcal E}}

\def\ZZ{{\mathbb Z}}

\nc{\sind}{\sigma{\rm -ind}}
\begin{document}


\title[Twisted K-homology theory, twisted $Ext$-theory]
{Twisted $K$-homology theory, twisted $Ext$-theory}
\author{V. Mathai}
\address{Department of Mathematics, MIT,
Cambridge, Mass 02139, USA and 
University of
Adelaide, Adelaide 5005, Australia}
\author{I.M. Singer}
\address{Department of Mathematics, MIT, Cambridge MA 02139}
\thanks{{V.M.  acknowledges support from the Clay 
Mathematical Institute. I.M.S. acknowledges support from
DOE contract \# DE-FG02-88ER25066}}
\begin{abstract}
These are notes on twisted $K$-homology 
theory and twisted $Ext$-theory from the $C^*$-algebra
viewpoint, part of a series of talks on ``$C^*$-algebras, 
noncommutative geometry and $K$-theory'', primarily for physicists.
\end{abstract}

\maketitle
\vspace{-.175in}

\section*{Index of notation}

\noindent $\bullet$ $\cK$ is the algebra of compact operators on a
(separable, infinite dimensional) Hilbert space
$\mathcal H$.\\
$\bullet$ $Aut(\cK)$ is the group of automorphisms of $\cK$.\\
$\bullet$ $PU = PU(\cH)= U(\cH)/U(1)$ is the group of projective unitary automorphisms
of the Hilbert space $\cH$. We will often identify 
$PU$ with $Aut(\cK)$ using the canonical isomorphism between
these groups.\\
$\bullet$ $\cB$ will often denote the algebra $C_0(X, {\mathcal E}_{H})$
of sections, vanishing at infinity, of the unique  locally trivial
bundle ${\mathcal E}_{H}$ over $X$ with fibre
$\cK$ and structure group $Aut(\cK)$ whose Dixmier-Douady invariant (see introduction),
$\delta({\mathcal E}_{H}) = [H] \in H^3(X, \mathbb Z)$.\\
$\bullet$ $\cA$ will often denote an algebra obtained as an 
extension of $\cB$ by $\cK$.\\
$\bullet$ $P_H$ is the unique principal $Aut(\cK)$-bundle over $X$ whose
Dixmier-Douady invariant,
$\delta(P_H) = [H] \in H^3(X, \mathbb Z)$.\\
$\bullet$ $\cF$ denotes the space of all Fredholm operators on a Hilbert space
$\cH$\\
$\bullet$ $\cQ=\cQ(\cH)$ denotes the Calkin algebra, that is $\cQ = B(\cH)/\cK$
where $B(\cH)$ is the algebra of all bounded operators on $\cH$.\\

\section{Introduction}

Long ago, Dixmier and Douady \cite{DD} observed that the algebra bundles with fibre $\cK$  
over a locally compact space $X$ were classified by $[H] \in H^3(X, \mathbb Z)$
(because $Aut(\cK) \cong PU$ and $\pi_j(PU) = 0$
for $j\ne 2$ but $\pi_2(PU) \cong \mathbb Z$).

Let $P_H$ be a principal bundle over $X$ with fibre $PU$ and 
$\delta(P_H) = [H] \in H^3(X, \mathbb Z)$. (Transgression etc). Let $\E_H$
be the bundle $P_H\times_G \cK$ and $\cF_H$ the bundle
$P_H\times_G \cF$, where $G= PU= Aut(\cK)$,
also acts on $\cK$ by conjugation. Let $C_0(X, \cE_H)$ be the continuous sections
of $\E_H$ vanishing at infinity. Rosenberg \cite{Ros} defined twisted $K$-theory 
(section 2) and showed that $\cF_H$ is the classifying space 
for twisted $K^0$, an extension of the 
well known theorem for $H=0$, i.e., $\cF$ is the classifying space for ordinary
$K^0$. 

In recent times, twisted $K$-theory has entered string/M-theory when the 
$H$-field of the Neveu-Schwarz sector is turned on. See Bouwknegt-Mathai \cite{BM}, 
Witten \cite{Wi},
for the linking of the physics with the $C^*$-algebras, and the references there
to twisted $K$-theory. There has been considerable speculation about the
role for $K$-homology (even for $C^*$-algebras) in $D$-brane theory. See
Harvey-Moore \cite{HaMo} and the references there.

Once one is in the twisted category, one asks whether $K$-homology, $Ext$-theory, and 
Fredholm modules can similarly be twisted and with the usual relations between them.
This is indeed the case, and is really part of Kasparov's $KK$-theory, a 
prediction made to the second author by R.G. Douglas.

These notes are a short exposition of $K$-homology, $Ext$-theory, and Fredholm modules 
in the twisted category from the $C^*$-algebra viewpoint. They
 are meant primarily for physicists. One can also develop a twisted theory for any
generalized cohomology theory. How to do so was explained to us by M.J. Hopkins.

\section{Twisted $K$-theory and noncommutative geometry}

Let $X$ be a locally compact, Hausdorff space with a countable basis of
open sets, for example a smooth manifold. Let $[H] \in H^3(X, \ZZ)$. Then
the {\em twisted $K$-theory} was defined by Rosenberg  as
\begin{equation}
K^j(X, [H]) = K_j(C_0(X, {\mathcal E}_{H}))\qquad j=0,1,
\end{equation}
where ${\mathcal E}_{H}$ is the unique  locally trivial
bundle over $X$ with fibre
$\cK$ and structure group $Aut(\cK)$ whose Dixmier-Douady invariant,
$\delta({\mathcal E}_{H}) = [H]$,
and $K_\bullet(C_0(X, {\mathcal E}_{H}))$ denotes the
topological $K$-theory of the
$C^*$-algebra of continuous sections of ${\mathcal E}_{H}$
that vanish at infinity. See \cite{Black} or \cite{Singer} 
for the definition of the topological $K$-theory of 
$C^*$-algebras.
Notice that when $H=0$, then ${\mathcal E}_{H} = X\times \cK;$ therefore
$C_0(X, {\mathcal E}_{H}) = C_0(X)\otimes \cK$ and
by Morita invariance of $K$-theory (cf. \cite{Black} or \cite{Singer}), 
the twisted $K$-theory of
$X$ coincides with  the standard $K$-theory of $X$ in this case.
Elements of $K^0(X, [H])$ are called (virtual) gauge-bundles in the
physics literature, but we will call these twisted bundles in these notes.

In \cite{Ros}, it is shown that
when $X$ is compact one has
\begin{equation}
\begin{array}{lcl}
K^0(X, [H]) &=& [P_H, \cF]^{PU}\\
K^1(X, [H]) &=& [P_H, U(\cK^+)]^{Aut(\cK)}\,
\end{array}
\end{equation}
where  $U(\cK^+)$ is the group of unitaries in the unitalization of
$\cK$, 
$$
U(\cK^+) = \left\{u\in U(\cH): u-1\in \cK\right\}.
$$

\section{The twisted Ext group and twisted K-homology}

In this section we will give a
brief review of the twisted $Ext$ group, which 
can be considered as a specialization of the general 
$Ext$-theory \cite{PPV}, \cite{Kas}.

Consider noncommutative $C^*$ algebras $\CA$ which
fit into the short exact sequence:
\begin{equation}\label{exact}
0 \rightarrow \cK \rightarrow \CA {\stackrel\beta\rightarrow} 
\cB
\rightarrow 0
\end{equation}
where $\cB = C_0(X, {\mathcal E}_{[H]})$, 
for some fixed space $X$ and NS field $H$. 
In \cite{PPV}, \cite{Kas} extensions of the form (\ref{exact}) for
general nuclear $C^*$-algebras $\cB$ were investigated. 
We shall restrict ourselves to the special case when 
$\cB = C_0(X, {\mathcal E}_{[H]})$.
To any such extension one can associate 
the Busby invariant, which is a homomorphism
\begin{equation}
\tau: C_0(X, {\mathcal E}_{[H]}) \to Q(\CH)
\end{equation}
defined as follows. 
For any $s\in C_0(X, {\mathcal E}_{[H]})$ we choose an
operator $T_s\in \CA$ such that $\beta(T_s) = s$, and define $\tau$ by:
$\tau(s) = \pi(T_s)$ where $\pi: B(\CH) \to Q(\CH)$ is the
projection.   $\tau$ is a homomorphism
because $T_{s_1} T_{s_2} - T_{s_1 s_2}$ is a compact
operator.  
Conversely, given
a homomorphism $\tau: C_0(X, {\mathcal E}_{[H]}) \to Q(\CH)$ one can
form an extension \ref{exact} as follows,
\begin{equation}\label{exact2}
0 \rightarrow \CK \rightarrow \CA' ~ \rightarrow~ C_0(X, {\mathcal E}_{[H]}) \rightarrow 0
\end{equation}
where the algebra $\CA'$ is defined as
\begin{equation}
\CA' = \{ (A, f) : \pi(A) = \tau(f) \} \subset B(\CH) \oplus C_0(X, {\mathcal E}_{[H]}).
\end{equation}
Moreover, (\ref{exact}) is unitarily equivalent to (\ref{exact2}) in the sense
that we now describe.

Two extensions (\ref{exact}) are unitarily equivalent if there is
a unitary operator $U$ on $\CH$ such that the
Busby invariants are related by $\tau_2(s) = \pi(U) \tau_1(s) \pi(U)^*$.
Let $\; {\bf Ext}(X, H)$ denote the set of unitary equivalence
classes of extensions of $C_0(X, {\mathcal E}_{[H]})$ by $\CK$.
A direct sum operation on ${\bf Ext}(X, [H])$ can then be
defined by taking the extension corresponding
to the Busby invariant
\begin{equation}\label{sum}
\tau_1 \oplus \tau_2: C_0(X, {\mathcal E}_{[H]}) \rightarrow Q(\CH) \oplus Q(\CH) \rightarrow
Q(\CH \oplus \CH) \cong Q(\CH).
\end{equation}
Then (\ref{sum}) defines a semigroup operation on ${\bf 
Ext}(X, H)$. Trivial extensions are those for which the Busby invariant
lifts to $B(\CH)$. Equivalently, they are extensions such that
  the sequence (\ref{exact}) splits.
Define the {\em twisted $Ext$ group} $Ext(X, [H])$ as being the quotient of
${\bf Ext}(X, H)$ by the trivial extensions.
It is shown in \cite{PPV}, \cite{Kas} that
 every extension has an inverse up to the addition of a trivial
extension, so that $Ext( X, [H])$
is an abelian group. It is clear that $Ext( X, [H])$ depends 
only on the cohomology class of $[H]$, since $C_0(X, {\mathcal E}_{[H]})$
and $C_0(X, {\mathcal E}_{[H']})$ are isomorphic whenever $[H'] = [H]$.

There is a pairing of $Ext( X, [H])$ and $K^1(X, [H]) $ defined as follows.
(See \cite{BM}, \cite{DK}, \cite{Ka}, \cite{Ros}, \cite{Wi} for a discussion  
of twisted $K$-theory, $K^\bullet(X, [H])$).
An extension of the form (\ref{exact}) is determined by its Busby invariant
$\tau : C_0(X, {\mathcal E}_{[H]}) \to Q(\CH)$, which induces homomorphisms
$\widetilde\tau : M_n(C_0(X, {\mathcal E}_{[H]})^+) \to M_n(Q(\CH)) \cong Q(\CH)$
for each $n\in \mathbb N$.
If $u$ is a unitary in $M_n(C_0(X, {\mathcal E}_{[H]})^+)$, define the pairing
\begin{equation}
\begin{array}{rcl}
Ext( X, [H]) \times K^1(X, [H]) & \to & \mathbb Z\cr
(\tau, u) & \to &{\rm Index}(\widetilde\tau(u)).
\end{array}
\end{equation}
In particular, each element $\tau \in Ext( X, [H])$ defines a homomorphism
$\tau_* : K^1(X, [H])  \to \mathbb Z$. If $\tau_* = 0$, then the six term
exact sequence in K-theory (cf. section 4) corresponding to the extension (\ref{exact})
reduces to the short exact sequence
\begin{equation}
0 \rightarrow \mathbb Z = K_0(\cK) \rightarrow K_0(\CA) {\rightarrow} 
K^0(X, [H])
\rightarrow 0
\end{equation}
and therefore  defines an element of $Ext_{\mathbb Z}^1(K^0(X, [H]), \mathbb Z)$
in homological agebra. It can 
be shown \cite{RS} that the converse is also true, that is, one has the universal 
coefficient theorem:
\begin{equation}\label{universal}
0\to Ext_{\mathbb Z}^1(K^0(X, [H]), \mathbb Z) \to Ext( X, [H]) \to 
Hom(K^1(X, [H]), \mathbb Z) \to 0.
\end{equation}

This justifies
 the definition of 
the twisted $K$-homology as being
\begin{equation}
\begin{array}{lcl}\label{khomology}
K_1(X, [H]) & =& Ext( X, [H]) \cr
K_0(X, [H]) & =& Ext( X \times \mathbb R, p_1^*[H]) 
\end{array}
\end{equation}
where $p_1: X\times \mathbb R \to X$ denotes projection onto the 
first factor.

One deduces from the definition (\ref{khomology})  that the 
universal coefficient exact sequence (\ref{universal}) can be rewritten as
\begin{equation}\label{universal2}
0\to Ext_{\mathbb Z}^1(K^{\bullet +1}(X, [H]), \mathbb Z) \to K_\bullet( X, [H]) \to 
Hom(K^{\bullet}(X, [H]), \mathbb Z) \to 0.
\end{equation}

\section{Properties of the twisted $Ext$ groups and twisted $K$-homology}

\noindent 1) {\bf Bott Periodicity:} Let $[H] \in H^3(X, \mathbb Z)$. Then one has
the Bott periodicity theorem for the $C^*$-algebra $C_0(X, \cE_H)$, 
\begin{equation}
Ext( X \times {\mathbb R}^2, p_1^*[H]) \cong Ext( X, [H])
\end{equation}
For details on Bott periodicity for $C^*$-algebras,  cf. \cite{Black} or 
\cite{Singer}. 

This shows that if we  define $K_j(X, [H]) = Ext( X \times {\mathbb R}^j, p_1^*[H]) $ for 
$j\in \mathbb N$, then there are at most only two distinct 
groups in this list, $K_1(X, [H])$ and
$K_0(X, [H])$. 

\noindent 2) {\bf Six term exact sequence:} Given a short exact
sequence
$$
0\to C_0(X, {\mathcal E}_{[H]})\to \cB \to \cJ  \to 0
$$
there is a six term exact sequence of twisted $Ext$ groups, 
which is obtained using Bott periodicity,
$$\begin{CD}
K^0(\cJ) @>>>  K^0(\cB) @>>> K_0(X, [H]) \\
\delta @AAA  &&  \delta @VVV \\
K_1(X, [H])  @<F_*<<  K^1(\cB) @<\Delta<< K^1(\cJ)
\end{CD}$$

This enables us to compute the twisted $Ext$ groups at least in 
some examples.
Consider the case when $X=S^3$ and $[H]$ is the
class  of the volume form on $S^3$ and $N$ is a positive integer. Then
we will compute $K_\bullet(S^3, N[H])$. Note that this can also be done 
using the universal coefficient theorem.

We consider the open cover of $S^3$ given by the upper and lower
hemispheres, $\{\cU_1, \cU_2\}$, where $\cU_1\cap\cU_2 = S^2$. Then
representatives of $K^1(S^3, N [H]) $ are pairs of maps $(f_1, f_2)$,
where
$$
f_i: \cU_i \to U(\cK^+)
$$
such that on the overlap $\cU_1\cap\cU_2 = S^2$, one has
\begin{equation}
f_1 = p_{N[H]} f_2,
\end{equation}
where $ p_{N[H]}$ denotes the transition functions of the bundle 
$\cK$-algebra bundle $\cE_{N[H]}$ with Dixmier-Douady 
invariant $N [H] \in H^3(S^3, \mathbb Z)$.
Now the $C^*$-algebra of continuous sections of the bundle $\cE_{N[H]}$,
$C(X, \cE_{N[H]})$
can be represented by pairs of continuous functions $(h_1, h_2)$, where
$$
h_i : \cU_i \to \cK\qquad i=1,2
$$
and satisfying on the overlap $\cU_1\cap\cU_2 = S^2$
$$
h_1 = p_{N[H]} h_2.
$$
Therefore there is a short exact sequence
\begin{equation}\label{seq}
0\to C(S^3, \cE_{N[H]}) \stackrel{F}\to C(\cU_1)\otimes \cK\oplus C(\cU_2)\otimes \cK
\stackrel{G}\to C(S^2)\otimes \cK\to 0
\end{equation}
where
$$
F(h_1, h_2) = h_1 \oplus h_2, \qquad G(q_1 \oplus q_2) = q_1|_{S^2}
- p_{N[H]} \left(q_2|_{S^2}\right).
$$
The six term exact sequence in $Ext$-theory associated to the
short exact sequence $(\ref{seq})$ is,
$$\begin{CD}
K_0(S^2) @>G^*>>  K_0(\cU_1)\oplus K_0(\cU_2) @>F^*>> K_0(S^3, N[H]) \\
\delta @AAA  &&  \delta @VVV \\
K_1(S^3, N[H])  @<F^*<<  K_1(\cU_1)\oplus K_1(\cU_2) @<G^*<< K_1(S^2)
\end{CD}$$
Since $0 = K_1(\cU_1) \oplus K_1(\cU_2) = K_1(S^2)$, this six term exact
sequence collapses into the exact sequence
\begin{equation}
0 \to  K_1(S^3, N [H])  \stackrel{\delta}\longrightarrow K_0(S^2)
 \stackrel{G^*}\longrightarrow K_0(\cU_1) \oplus
K_0(\cU_2)
\stackrel{F^*}\longrightarrow  K_0(S^3, N [H]) \to 0
\end{equation}
On analyzing the map $G^*$ explicitly, we see that if $N\ne 0$, then
$Ker(G^*) = 0 = K_1(S^3, N [H])$ and that $Coker(G^*) = \ZZ_N \cong  K_0(S^3, N [H])$.
See \cite{Ros} and \cite{BM} for related computations.

\noindent 3) {\bf When $H$ is torsion:} In this case, there is an argument in 
\cite{Gr},\cite{Wi} which  shows that the free part of $K^0(X, [H])$ is 
isomorphic to the free part of $K^0(X)$. Therefore $Hom(K^0(X, [H]), \mathbb Z)
\cong Hom(K^0(X), \mathbb Z)$, and by the universal coefficient theorem, 
we see that the free part of the twisted $K$-homology $K_0(X, [H])$ is 
isomorphic to the free part of $K_0(X)$ in this case, and in particular,
$K_0(X, [H]) \otimes \mathbb Q \cong K_0(X) \otimes \mathbb Q$.

\noindent 4) {\bf The fundamental class when $H=0$:} 
A noteworthy case, to be reviewed later in these
talks, is the extension
\begin{equation}\label{pdo}
0\to \cK \to \cA \to C(S^*X)\to 0
\end{equation}
where $\cA$ is the closure in the norm topology of the algebra of singular
integral operators (pseudodifferential operators of order zero) and 
$S^*X$ is the sphere bundle of the cotangent bundle of a smooth manifold $X$.
The extension (\ref{pdo})  does not 
split, and is called the fundamental class in $Ext(S^*X)$, cf. \cite{Kas}, \cite{BD}.

\section{Description of twisted $K$-homology in terms of Fredholm operators}

We recall here the definition of $KK^1(\cB, \mathbb C)$, 
where $\cB = C_0(X, {\mathcal E}_{[H]})$ and $\mathbb C$ is the algebra of complex numbers,
a very special case of Kasparov's $KK(\cB, \cD)$ theory for general
$C^*$-algebras $\cD$. This will 
provide us with a Fredholm module picture for twisted $Ext$-theory. A Fredholm
module is a  triple $(\cH, \phi, F)$, where,
\begin{itemize}
\item $\cH$ is a separable Hilbert space;
\item $\phi : C_0(X, {\mathcal E}_{[H]}) \to B(\cH)$ is a $*$-homomorphism;
\item $F$  is self-adjoint and satisfies:
 $(F^2 - 1)\phi(a) \in \cK,$
and  $[F, \phi(a)]  \in \cK$ for all $a\in \cB$. 
\end{itemize}
Let ${\rm E}_1(\cB)$ denote the set of all Fredholm modules
over $\cB$.
 Let ${\rm D}_1(\cB)$ denote the subset of Fredholm modules satisfying
$(F^2 - 1)\phi(a) = 0 = [F, \phi(a)] $. They are called degenerate Fredholm modules.

The direct sum of two Fredholm modules is 
again a Fredholm module. Moreover, the direct sum of degenerate Fredholm modules 
is again a degenerate Fredholm module. Two Fredholm 
modules $(\cH_i, \phi_i, F_i),\;i=0,1$
are said to be unitarily equivalent if there is a unitary in $B(\cH_0, \cH_1)$
intertwining the $\phi_i$ and the $F_i$. 

Define an equivalence relation $\sim$ on ${\rm E}(\cB)$ 
generated by unitary equivalence,
addition of degenerate elements and `compact perturbations' of $(\cH, \phi, F)$. Here 
a Fredholm module $(\cH, \phi, F')$ is said to be a 
compact perturbation of $(\cH, \phi, F)$ if $(F-F')\phi(a) \in \cK$ for all
$a\in \cB$. 

Then $KK^1(\cB, \mathbb C)$ is the set of equivalence classes of ${\rm E}_1(\cB)$
under the equivalence relation $\sim$. 

Given a Fredholm module $(\cH, \phi, F)$, we will define 
a $\cK$ extension of $\cB$ of the form 
(\ref{exact}) as follows. Observe that $P = 1/2 F + 1/2$ is a projection modulo 
$\cK$. Define the Busby map $\tau$ by $\tau(a) = \pi(P \phi(a) P)$ for all 
$a\in \cB$, where $\pi : B(\cH) \to Q(\cH)$ is the projection. 
Then $\tau$ gives the desired
$\cK$ extension of $\cB$ of the form (\ref{exact}). The Busby map corresponding 
to $1-P$ is an inverse for $\tau$, and 
we have a well defined map 
$$
KK^1(\cB, \mathbb C) \to Ext(X, [H]) = K_1(X, [H]).
$$
It is not too hard to show that this map is an isomorphism
$$
KK^1(\cB, \mathbb C) \cong Ext(X, [H]) = K_1(X, [H]).
$$
This gives a Fredholm module description of twisted $Ext$-theory, or equivalently
of twisted $K$-homology theory.

There is also a $\mathbb Z_2$-graded Fredholm module description of the 
twisted $K$-homology group
$K_0(X, [H])$, which we will now discuss. A $\mathbb Z_2$-graded Fredholm module is a 
triple $(\cH, \phi, F)$, where $\cH$ is a separable $\mathbb Z_2$-graded Hilbert space,
$\phi : C_0(X, {\mathcal E}_{[H]}) \to B(\cH)$ is a $*$-homomorphism which 
is of even degree, $F$ is an odd degree self-adjoint operator on $\cH$ and satisfies 
$(F^2 - 1)\phi(a) \in \cK,\;\; [F, \phi(a)] 
\in \cK$ for all $a\in \cB$. Let ${\rm E}_0(\cB)$ denote the set of all 
$\mathbb Z_2$-graded Fredholm modules
over $\cB$. Let ${\rm D}_0(\cB)$ denote the subset of $\mathbb Z_2$-graded 
Fredholm modules satisfying
$(F^2 - 1)\phi(a) = 0 = [F, \phi(a)] $. They are called degenerate 
$\mathbb Z_2$-graded Fredholm modules.

The direct sum of two $\mathbb Z_2$-graded Fredholm modules is 
again a $\mathbb Z_2$-graded Fredholm module, with respect to 
the total $\mathbb Z_2$-grading.
 Moreover, the direct sum of degenerate $\mathbb Z_2$-graded Fredholm modules 
is again a degenerate $\mathbb Z_2$-graded Fredholm module. Two 
$\mathbb Z_2$-graded Fredholm modules $(\cH_i, \phi_i, F_i),\;i=0,1$
are said to be unitarily equivalent if there is a unitary in $B(\cH_0, \cH_1)$
intertwining the $\phi_i$ and the $F_i$. 

Define an equivalence relation $\sim$ on ${\rm E}_0(\cB)$ 
generated by unitary equivalence,
addition of degenerate elements and `compact perturbations' 
of $(\cH, \phi, F)$. Here 
a $\mathbb Z_2$-graded Fredholm module $(\cH, \phi, F')$ is said to be a 
compact perturbation of $(\cH, \phi, F)$ if $(F-F')\phi(a) \in \cK$ for all
$a\in \cB$. 

Then $KK^0(\cB, \mathbb C)$ is the set of equivalence classes of ${\rm E}_0(\cB)$
under the equivalence relation $\sim$. 
It follows from the discussion above and Bott periodicity that\\
$
KK^0(\cB, \mathbb C) = KK^1(\cB\otimes C_0(\mathbb R), \mathbb C) 
\cong Ext(X\times \mathbb R, p_1^*[H]) = K_0(X, [H]).
$

\section{Topological $K$-homology}

We now give a Baum-Douglas type description of 
twisted $K$-homology, called topological twisted 
$K$-homology. The basic objects are 
twisted $K$-cycles. 
A {\em twisted} $K${\em-cycle} on a 
topological space is a triple $(M, E, \phi)$, where $M$ is a compact 
Spin$^{\mathbb C}$ manifold, $E\to M$ is a  twisted bundle on $M$, and 
$\phi : M \to X$ is a continuous map. Two twisted $K$-cycles $(M, E, \phi)$ and 
$(M', E', \phi')$ are said to be 
{\em isomorphic} if  there is a diffeomorphism $h: M \to M'$ such that 
$h^*(E') \cong E$ and $h^*\phi' = \phi$. Let $\Pi(X, H)$ denote the 
collection of all twisted $K$-cycles on $X$. 

\noindent $\bullet$ {\em Bordism}: $(M_i, E_i, \phi_i) \in
\Pi(X, H)$,
$i=0,1$ are said to be  {\em bordant} if there is a
triple $(W, E, \phi)$ where $W$ is a compact 
Spin$^{\mathbb C}$ manifold with boundary $\partial W$,
$E$ is a twisted bundle over $W$ and $\phi : W
\to X$ is a continuous map, such that
$(\partial W, E\big|_{\partial W}, \phi\big|_{\partial W})$ is isomorphic
to the disjoint union $(M_0, E_0, \phi_0) \cup (-M_1, E_1, \phi_1)$. Here
$-M_1$ denotes $M_1$ with the reversed Spin$^{\mathbb C}$ structure.

\noindent $\bullet$ {\em Direct sum}: Suppose that  $(M,
E, \phi) \in \Pi(X, H)$ and that
$E=E_0\oplus E_1$. Then $(M, E, \phi)$ is isomorphic to $(M, E_0, \phi)
\cup (M, E_1, \phi)$.

\noindent $\bullet$ {\em Twisted bundle modification}: Let
$(M, E, \phi) \in \Pi(X, H)$ and
${\bf H}$ be an even dimensional Spin$^{\mathbb C}$ 
vector bundle over M. Let $\widehat M = S({\bf H}\oplus 1)$ denote the sphere 
bundle of ${\bf H}\oplus 1$. Then $\widehat M$ is canonically a  Spin$^{\mathbb C}$ 
manifold. Let ${\mathcal S}$ denote the bundle of spinors on ${\bf H}$. Since 
${\bf H}$ is even dimensional, ${\mathcal S}$ is ${\mathbb Z}_2$-graded, 
$$
{\mathcal S} = {\mathcal S}^+ \oplus {\mathcal S}^-
$$
into bundles of $1/2$-spinors on $M$. Define $\widehat E = \pi^*(
{\mathcal S}^{+*} \otimes E)$, where $\pi : \widehat M \to M$ is the
projection. Finally,  $\widehat \phi = \pi^*\phi$. Then $(\widehat M,
\widehat E, \widehat\phi) \in \Pi(X, H)$ is said to be obtained from 
$(M, E, \phi)$ and ${\bf H}$ by {\em twisted bundle modification}. 

Let $\;\sim\;$ denote the equivalence relation on
$\Pi(X, H)$ generated by the operations of bordism, direct
sum and twisted bundle modification. Notice that
$\;\sim\;$ preserves the parity of the dimension of 
the twisted $K$-cycle. Let $K_0^t(X, [H])$ denote the quotient
$\Pi^{even}(X, H)/\sim$, where 
$\Pi^{even}(X, H)$ denotes the set of all even dimensional 
twisted $K$-cycles in
$\Pi(X, H)$, and let $K_1^t(X, [H])$ denote the quotient 
$\Pi^{odd}(X, H)/\sim$, where 
$\Pi^{odd}(X, H)$ denotes the set of all odd dimensional 
twisted $K$-cycles in
$\Pi(X, H)$. Then it is possible to show as in \cite{BD} 
that  $K_j^t(X, [H]) \cong K_j(X, [H]), \; j=0, 1,$ 
providing a topological 
description of twisted $K$-homology.


\begin{thebibliography}{CHMM}

\bibitem[BD]{BD} P. Baum and R. Douglas,  $K$ homology and index theory, 
Operator algebras and applications,
Part I (Kingston, Ont., 1980), pp. 117-173, Proc. Sympos. Pure Math., 
38, Amer. Math. Soc., Providence, R.I., 1982.

\bibitem[Black]{Black} B. Blackadar, K-theory for operator algebras,
MSRI publications, vol 5, Cambridge University Press, 1986.


\bibitem[BM]{BM} P.~Bouwknegt and V.~Mathai, {\em D-Branes, B-Fields 
and twisted
K-theory},  Journal of High Energy Physics, {\bf 03} (2000) 007 (11
pages); {\em ibid.}, in preparation.


\bibitem[DD]{DD} J.~Dixmier and A.~Douady, {\em Champs continues d'espaces
hilbertiens at de $C^*$-alg\`ebres}, Bull.\ Soc.\ Math.\ France\ {\bf 91}
(1963) 227--284.

\bibitem[DK]{DK} P.~Donovan and M.~Karoubi, {\em Graded Brauer
groups and $K$-theory with local coefficients}, Inst.\ Hautes
\'Etudes Sci.\ Publ.\ Math., {\bf 38} (1970) 5--25.

\bibitem[Gr]{Gr} A.~Grothendieck, {\em Le groupe de Brauer, I, II,
III. 1968 Dix Expos\'es sur la Cohomologie des Sch\'emas}, pp. 46--188,
(North-Holland, Amsterdam; Masson, Paris).

\bibitem[HaMo]{HaMo} J. Harvey, G. Moore,
Noncommutative Tachyons and K-Theory,
[{\tt hep-th/0009030}]

\bibitem[Ka]{Ka} A. Kapustin, {\em $D$-branes in a topologically nontrivial
$B$-field}, [{\tt hep-th/9909089}].

\bibitem[Kas]{Kas} G. Kasparov, Equivariant $KK$-theory and the Novikov
conjecture, {\em Inv.\ Math.} {\bf 91} (1988), 147-201.
{\em ibid.}, Topological invariants of elliptic operators, I. 
$K$-homology. (Russian) Math.
USSR-Izv. {\bf 9} (1975), no. 4, 751-792.

\bibitem[PPV]{PPV} M. Pimsner, S. Popa, D. Voiculescu, 
 Homogeneous $C\sp{*} $-extensions of $C(X)\otimes K(H)$. II. 
 J. Operator Theory {\bf 4} (1980), no. 2, 211-249; 
{\em ibid}, Homogeneous $C\sp{*} $-extensions of 
$C(X)\otimes K(H)$. I. J. Operator Theory 
{\bf 1} (1979), no. 1, 55-108. 

\bibitem[Ros]{Ros} J.~Rosenberg, {\em Continuous trace algebras
from the bundle theoretic point of view}, Jour.\ Austr.\ Math.\ Soc.,
{\bf 47} (1989), 368--381; {\em ibid.}, {\em Homological invariants of
extensions of $C^*$-algebras}, Proceedings of Symposia in Pure Mathematics,
{\bf 38} (1982) 35-75.

\bibitem[RS]{RS} J. Rosenberg and C. Schochet, The Kunneth theorem and the 
universal coefficient theorem for Kasparov's generalized
$K$-functor, {\em Duke Math. J.} {\bf 55} (1987), no. 2, 431-474. 

\bibitem[Singer]{Singer} V. Mathai and I.M. Singer, Lectures on 
operator algebras, noncommutative
geometry and $K$-theory (primarily for physicists), fall 2000, 
MIT (in preparation). 

\bibitem[Wi]{Wi} E.~Witten, {\em $D$-branes and $K$-theory},
JHEP {\bf 12} (1998) 019, [{\tt hep-th/9810188}]; {\em ibid.},
Overview Of K-Theory Applied To Strings, [{\tt hep-th/0007175}].

\end{thebibliography}
\end{document}